\documentclass[conference]{IEEEtran}
\IEEEoverridecommandlockouts
\usepackage{cite}

\usepackage{amsmath,amssymb,amsfonts}
\usepackage{enumerate}
\usepackage{algorithmic}
\usepackage{graphicx}
\usepackage{textcomp}
\usepackage{color,xcolor}
\usepackage{subfigure}
\usepackage{multirow,tabularx}
\usepackage{float}
\usepackage{subfigure}
\usepackage{CJK}

\usepackage[linesnumbered,ruled,vlined]{algorithm2e}
\usepackage{url}

\def\BibTeX{{\rm B\kern-.05em{\sc i\kern-.025em b}\kern-.08em
		T\kern-.1667em\lower.7ex\hbox{E}\kern-.125emX}}
\begin{document}
	
	\title{An Ordered-Reliability-Bits Chase Decoding Algorithm for BCH Codes \\
	}
	
	\author{
		\IEEEauthorblockN{Wenwu Zhu, Min Zhu, and Baoming Bai}
		\IEEEauthorblockA{State Key Laboratory of ISN, Xidian University, Xi'an, China}
		\IEEEauthorblockA{24011211309@stu.xidian.edu.cn, mzhu@xidian.edu.cn, bmbai@mail.xidian.edu.cn}
	}
	\maketitle
	
\begin{abstract}
In this paper, we propose a low-complexity ordered-reliability-bits Chase (ORB-Chase) decoding algorithm for BCH codes. The proposed algorithm differs from the traditional Chase algorithm in two key aspects. First, it employs the logical weight as a metric to generate test error patterns (TEPs). Second, it introduces an integer-based early termination criterion that ensures computation can stop at the earliest possible stage if the maximum-likelihood codeword is identified, thereby minimizing unnecessary computational effort. Simulation results for (127, 113, 5) BCH codes and (256, 239, 6) eBCH codes demonstrate that the ORB-Chase algorithm achieves near-ML performance with significantly fewer test patterns compared to the Chase algorithm. Moreover, the average number of Berlekamp-Massey (BM) decoding calls decreases rapidly as  $E_b / N_0$ increases, achieving a reduction of up to $98.1\%$ compared to the Chase algorithm at the same BLER performance.
\end{abstract}


\section{Introduction}

In recent years, short block-length channel coding techniques have attracted significant attention due to their critical role in scenarios such as ultra-reliable low-latency communications (URLLC) and optical transport networks (OTN) \cite{miao2024}\cite{Mahmood2023}. Short codes can also be used as component codes in more complex forward error correction (FEC) code structures, such as staircase codes \cite{Smith2012}, braided codes \cite{Truhachev2003}, and zipper codes \cite{Sukmadji2022}.

Bose-Chaudhuri-Hocquenghem (BCH) codes \cite{Berlekamp1968}, with their strong algebraic structure and error correction capability, especially their large minimum distance, have become the most commonly used short codes. Benefiting from a well-established algebraic structure, BCH codes possess efficient hard-decision decoding algorithms, such as the Berlekamp-Massey (BM) algorithm \cite{massey1969}. However, hard-decision decoding algorithms are limited by the inherent error correction capability of the code, making it difficult to fully achieve their potential performance. Therefore, the study of efficient and high-performance soft-decision decoding algorithms holds significant theoretical and practical value.

One class of soft-decision decoding algorithms is Guessing Random Additive Noise Decoding (GRAND) \cite{Duffy2019}. During the decoding process, a sequence of noise sequences is generated in a descending order of likelihood (from the most probable to the least probable), terminating when a valid codeword is successfully decoded. Depending on the order in which noise sequences are generated, multiple variants of GRAND can be implemented, such as soft GRAND (SGRAND) \cite{Solomon2020}, discretized soft GRAND \cite{Yuan2023}, ordered reliability bits GRAND \cite{Duffy2022}, and so on. There is also a GRAND optimized for BCH codes \cite{Wang2025}, which performs GRAND in the first stage and employs algebraic decoding in the second stage.

Another class of soft-decision decoding algorithms is the reliability-based decoding algorithm. They are mainly divided into ordered statistics decoding (OSD) algorithms \cite{Fossorier1995} and Chase algorithms \cite{Chase1972}, which are based on the most reliable basis (MRB) and the least reliable basis (LRB), respectively. Reliability-based decoding algorithms can control the decoding complexity by adjusting the codeword search range, thereby achieving good decoding performance with low algorithmic complexity. However, as performance requirements increase, the algorithmic complexity also increases dramatically. Therefore, reducing complexity while maintaining performance gains remains a pressing research challenge for this class of algorithms.

To address this problem, we propose a low-complexity ordered-reliability-bits Chase (ORB-Chase) decoding algorithm for BCH codes in this paper. By optimizing the generation method of test sequences and introducing an early termination condition, the efficiency of soft-decision decoding for BCH codes can be significantly improved. The rest of the paper is organized as follows. Section II briefly reviews the preliminaries of BCH codes and the conventional Chase decoding algorithm. Section III details the proposed ORB-Chase decoding algorithm. Numerical results are presented in Section IV and Section V concludes the paper.

\section{Preliminaries}
\subsection{BCH Codes}
As an important class of cyclic codes, BCH codes are capable of correcting multiple random errors. We use the triple $(n,k,d_{\min})$ to denote the parameters of a BCH code, where $n$ is the code length, $k$ is the number of information bits, and $d_{\min}$ is the minimum Hamming distance. The parity-check matrix of a \(t\)-error-correcting BCH code can be expressed as
\[
H =
\begin{bmatrix}
1 & \alpha & \alpha^2 & \cdots & \alpha^{n-1} \\
1 & \alpha^2 & \alpha^4 & \cdots & \alpha^{2(n-1)} \\
\vdots & \vdots & \vdots & \ddots & \vdots \\
1 & \alpha^{2t} & \alpha^{2\cdot 2t} & \cdots & \alpha^{2t(n-1)}
\end{bmatrix},
\]
where \(\alpha\) is a primitive element in \(GF(2^m)\). For \( d \leq 2t \), any \( d \) columns of the parity-check matrix \( H \) are linearly independent; therefore, the BCH code given by the null space of \( H \) has minimum distance at least $2t+1$.

An extended BCH (eBCH) code is obtained by adding one overall parity-check bit to the BCH code. That is, the modulo-2 sum of the codeword of length \(n\) is computed to obtain the final parity bit, making the Hamming weight of the code even. The eBCH codes reduce the probability of error correction at the cost of a small rate loss \cite{Hager2018}.
\subsection{Chase Decoding}
\begin{figure}\center	
	\includegraphics[width=8.8cm]{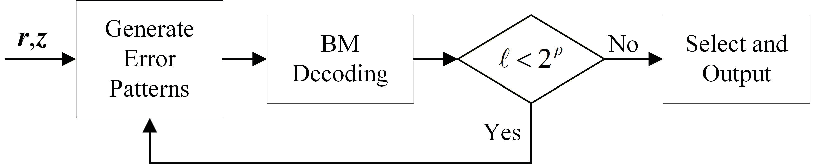}\\
	\caption{The diagram of the Chase decoding.}
	\label{chase}
\end{figure}

The Chase algorithm is a reliability-based decoding algorithm. Typically, the reliability of a codeword can be expressed by the log-likelihood ratio (LLR). When the received value is $\boldsymbol{r}$, the LLR of the \(i\)-th bit in the codeword $\boldsymbol{v}$ is calculated as
\begin{equation}
LLR(v_i) = \ln \left( \frac{P(v_i = 0 \mid r_i)}{P(v_i = 1 \mid r_i)} \right)
\end{equation}
where \(P(v_i = 0 \mid r_i)\) denotes the probability that \(v_i = 0\) given the received value \(r_i\), and \(P(v_i = 1 \mid r_i)\) denotes the probability that \(v_i = 1\) given \(r_i\). The absolute value of LLR reflects the confidence level of the corresponding bit; the larger the absolute value, the higher the reliability.
Under binary phase-shift keying (BPSK) modulation and additive white Gaussian noise (AWGN) channels, the received sequence $\boldsymbol{r}$ can be directly used to represent the LLR values of each bit. All discussions in this paper are based on this condition.

The schematic diagram of the Chase decoding is shown in Fig. \ref{chase}. By sorting the magnitudes of the received LLRs, Chase decoding generates \(2^p\) test patterns by flipping all \(2^p\) possible combinations of the \(p\) least reliable bits. Each pattern activates a hard-decision decoder, which produces a list of at most \(2^p\) codewords. The candidate codeword with the best soft-decision metric from the list is selected as the decoding output.

\section{Ordered Reliability Bits Based Chase Decoding Algorithm}
 To achieve high performance while reducing complexity, we propose an ordered reliability bits based Chase decoding algorithm in this section. This algorithm achieves near-optimal performance with only a small number of searches. The proposed algorithm differs from the Chase algorithm in two aspects. First, the logical weight is used as a metric to generate test error patterns (TEPs). Second, an integer-based early termination criterion is used to  determine whether the generated candidate codeword is the maximum-likelihood (ML) codeword. If the ML codeword is found, the decoding can be terminated.

\subsection{TEPs Based on Logical Weight}
In GRAND, error patterns are examined in a descending order of likelihood (from the most probable to the least probable). Therefore, the ordering of error patterns is central to different versions of GRAND. In the AWGN channel, the likelihood of a noise sequence is determined by the sum of the LLRs of the flipped bits. It is a non-trivial challenge to sort them precisely in various GRAND implementations.

Ordered reliability bits (ORB) technology has been proposed to simplify the process of the precise ordering. It has been observed that in the low-SNR range, ordered reliability values exhibit an approximately linear relationship with their ranks, with an intercept of zero \cite{Duffy2022}. Therefore, we can use the corresponding rank order to represent bit reliability, which leads to the baseline ORB. On this basis, the logical weight of a test pattern ${e^n}$ \cite{Duffy2022} can be defined as
\begin{equation}
	{w_L}({e^n}) = \sum\limits_{i = 1}^n i {e_i},
\end{equation}
where $i$ is the position of the flipped bit after sorting bits in ascending order of reliability (from low to high). That is, $i=1$ corresponds to the least reliable bit, and $i=n$  corresponds to the most reliable bit. Since the logical weight can approximate the likelihood of the noise sequence, TEPs can be generated in an ascending order of logical weights.

For a given logical weight, an integer partition pattern generator can be used to generate TEPs. This generator partitions a positive integer into distinct positive integers, with the number of parts ranging from small to large. This also implies that TEPs are sorted in the order of their Hamming weights when their logical weights are identical. For example, the integer 6 can be partitioned into $\left\{ 6 \right\},\{ 1,5\} ,\{ 2,4\} ,\{ 1,2,3\}$. By this generator, we can generate the required TEPs $e^{(i)} \in \mathbb{F}_2^n$  that satisfy
\begin{equation}
	{w_L}({e^{(0)}}) \le {w_L}({e^{(1)}}) \le  \cdots  \le {w_L}({e^{(\ell )}}).
\end{equation}

\subsection{Integer-based Early Termination Criterion}
The second feature of our algorithm is the use of a criterion to test whether a generated candidate codeword is the ML codeword. When the candidate codeword satisfies the criterion, the decoding process is terminated immediately without generating the entire candidate list. This early termination reduces computational complexity and decoding latency. In the following, we first review the Optimality Criterion in \cite{Kaneko1994}.

Let the transmitted signal $\boldsymbol{x} \in \mathbb{F}_2^n$  be modulated by BPSK and transmitted over an AWGN channel, resulting in the received sequence $\boldsymbol{r} \in \mathbb{R}^n$. Let $\boldsymbol{v}$ be a valid codeword of a BCH code, and let $\boldsymbol{z}$ be the hard-decision sequence of the received codeword. Let ${D_0}(\boldsymbol{v})$ and ${D_1}(\boldsymbol{v})$ denote the sets of indices where $\boldsymbol{v}$  agrees and disagrees with the hard-decision sequence $\boldsymbol{z}$, respectively:
\begin{equation}
{D_0}(\boldsymbol{v}) \buildrel \Delta \over = \left\{ {i:{v_i} = {z_i}\;{\rm{with}}\;0 \le i < n} \right\},
\end{equation}
\begin{equation}
\begin{array}{l}	
	\begin{aligned}
		D_1(\boldsymbol{v}) &\buildrel \Delta \over = \left\{ i: v_i \ne z_i \;\text{with}\; 0 \le i < n \right\} \\
		&= \left\{ 0,1, \ldots, n-1 \right\} \setminus D_0(\boldsymbol{v}).
	\end{aligned}	
\end{array}
\end{equation}
The cardinality of the set \({D_1}(\boldsymbol{v})\) is
\begin{equation}
	n(\boldsymbol{v}) = \left| {{D_1}(\boldsymbol{v})} \right|.
\end{equation}
Then the correlation discrepancy between $\boldsymbol{r}$ and $\boldsymbol{v}$ is given by
\begin{equation}
	\lambda (\boldsymbol{r},\boldsymbol{v}) = \sum\limits_{i \in {D_1}(\boldsymbol{v})} | {r_i}|.
\end{equation}

Sorting the indices in ${D_0}(\boldsymbol{v})$ according to the reliability measure of the received symbols, we have
\begin{equation}
{D_0}(\boldsymbol{v}) = \left\{ {{l_1},{l_2}, \ldots ,{l_{n - n(\boldsymbol{v})}}} \right\}
\end{equation}
such that for  $1 \le i < j \le n - n(\boldsymbol{v})$
\begin{equation}
	|{r_{{l_i}}}| < |{r_{{l_j}}}|.
\end{equation}
Let $D_0^{(j)}(\boldsymbol{v})$ denote the set of the first $j$ indices in the sorted set ${D_0}(\boldsymbol{v})$. That is,
\begin{equation}
	D_0^{(j)}(\boldsymbol{v}) = \left\{ {{l_1},{l_2}, \ldots ,{l_j}} \right\}.
\end{equation}
For \(j<1\), the set \(D_0^{(j)}(\boldsymbol{v})\) is defined as the empty set.

The codeword that minimizes $\lambda (\boldsymbol{r},\boldsymbol{v})$ is the ML codeword among all valid codewords $\boldsymbol{v}$. A tight lower bound on  $\lambda (\boldsymbol{r},\boldsymbol{v})$ can be used to determine whether a codeword is the ML codeword, leading to the following criterion \cite{Kaneko1994}.

\textit{Optimality Criterion}: Let $\rho={d_{\min }} - n(\boldsymbol{v}) $. If a codeword $\boldsymbol{v}$ satisfies
\begin{equation}\label{eq:criterion1}
	\lambda (\boldsymbol{r},\boldsymbol{v}) \le \sum\limits_{i \in D_0^{(\rho )}(\boldsymbol{v})} | {r_i}|,
\end{equation}
$\boldsymbol{v}$ is the ML codeword for the received sequence $\boldsymbol{r}$.

The logical weight is an integer form of reliability and can approximate the reliability. In \cite{Duffy2022}, a method of piecewise linear fitting of the reliability curve is also proposed. The one-line version of the method is as follows.

Sort the received sequence \(\boldsymbol{r}\) in non-decreasing order of reliability to obtain \(\tilde{\boldsymbol{r}}\). The reliability curve is fitted using a straight line passing through the points \((I_0, \tilde{r}_{I_0})\) and \((I_1, \tilde{r}_{I_1})\), where \( I_0 = 0 \) and \(I_1 = [n/2-1]\). The slope of the line is used as a quantization parameter, and it is given by
\begin{equation}
Q = \frac{\tilde{r}_{I_1} - \tilde{r}_{I_0}}{I_1 - I_0}.
\end{equation}
The quantized slope is set to 1, and the intercept is
\begin{equation}
J \equiv \left[ \frac{\tilde{r}_{I_0}}{Q} \right],
\end{equation}
where $[\cdot]$ is the rounding operation.

Let \(\varphi_i\) be the position of \(r_i\) in the sorted sequence \( \tilde{\boldsymbol{r}}\). Then the integer form of the reliability of \(r_i\) can be expressed as
\begin{equation}\label{eq:Integer}
\lambda_i = J + \varphi_i, \quad 0 \leq i < n.
\end{equation}

Motivated by the \textit{Optimality Criterion}, we replace the floating-point reliabilities with their corresponding integer reliability representation in \eqref{eq:Integer} and obtain an approximate criterion.

\textit{Integer-based Early Termination Criterion:} Let \(\rho = d_{\text{min}} - n(\boldsymbol{v})\). If a codeword \(\boldsymbol{v}\) satisfies
\begin{equation}\label{eq:criterion2}
\sum_{i \in D_1(\boldsymbol{v})} \lambda_i \leq \sum_{j \in D_0^{(\rho)}(\boldsymbol{v})} \lambda_j,
\end{equation}
then the decoding process can be terminated, and \(\boldsymbol{v}\) is regarded as the ML codeword for the received sequence \(\boldsymbol{r}\).

The proposed criterion approximates the \textit{Optimality Criterion} using integer reliabilities and therefore does not strictly ensure ML decoding. Nevertheless, simulations in Section IV indicate that its ML identification results are almost identical to those of the \textit{Optimality Criterion}.

\subsection{Ordered Reliability Bits Chase Decoding Algorithm}
\begin{figure}\center
	
	\includegraphics[width=8.8cm]{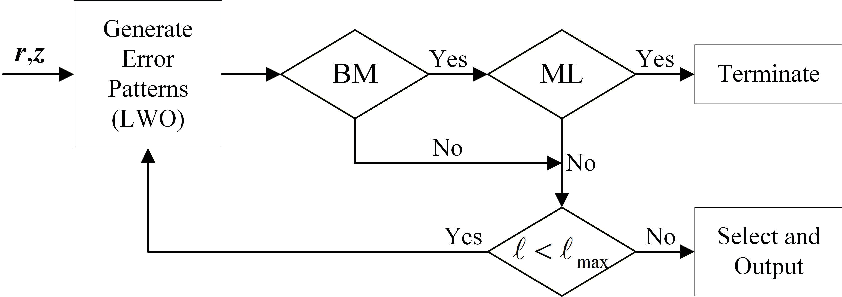}\\
	\caption{The diagram of the ORB-Chase decoding.}
	\label{decoding}
\end{figure}

In this subsection, we describe the proposed ordered reliability bits based Chase (ORB-Chase) decoding algorithm for BCH Codes in detail. The schematic diagram of the algorithm is shown in Fig. \ref{decoding}. The inputs of the decoder are the received sequence $\boldsymbol{r} \in \mathbb{R}^n$ and the hard-decision sequence  $\boldsymbol{z} \in \mathbb{F}_2^n$. The output is the information-bits sequence of the optimal codeword. The decoding process is detailed as follows.

\begin{itemize}
  \item \textbf{Generate Error Patterns and Test Sequences:} Based on the received sequence $\boldsymbol{r}$ and the ORB technique, TEPs \( e^{(i)} \in \mathbb{F}_2^n \), \( 0 \leq i \leq \ell_{\text{max}} - 1 \),
  are generated in logical weight order (LWO). It should be noted that when the logical weights are the same, the TEPs are generated in ascending order of Hamming weight.
  For each \( e^{(i)} \), the test sequence $\boldsymbol{a}^{(i)}$ is given by
  \[
  \boldsymbol{a}^{(i)} = \boldsymbol{z} \oplus e^{(i)}, \quad \text{for } 0 \leq i \leq \ell_{\text{max}} - 1.
  \]

  \item \textbf{BM Decoding:} After generating a test sequence, perform BM decoding on it. If decoding succeeds and a valid codeword ${\boldsymbol{v}^{(j)}}$ is obtained, put it into the set of candidate codewords ${\{\boldsymbol{v}^{(j)}\}}$.
  \item \textbf{ML Codeword Decision:} Once a valid codeword is obtained, determine whether it is the ML codeword according to the \textit{Integer-based Early Termination Criterion} \eqref{eq:criterion2}. If the criterion is satisfied, $\boldsymbol{v}^{(j)}$ is selected as the decoded codeword; terminate decoding and output the corresponding information bits $\boldsymbol{u} \in \mathbb{F}_2^k$.
  \item \textbf{Output:} If the process is not terminated early, after performing BM decoding \(\ell_{\text{max}}\) times, select the codeword with the smallest correlation discrepancy from the candidate codeword set ${\{\boldsymbol{v}^{(j)}\}}$ as the optimal codeword and output the corresponding information bits $\boldsymbol{u} \in \mathbb{F}_2^k$.
\end{itemize}

\begin{algorithm}
\DontPrintSemicolon
\SetKw{KwTo}{to}
\SetKw{KwBreak}{break}
\KwIn{Received signal $\boldsymbol{r} \in \mathbb{R}^n$, hard-decision sequence $\boldsymbol{z} \in \mathbb{F}_2^n$, maximum number of TEPs $\ell_{\max}$.}
\KwOut{Information data $\boldsymbol{u} \in \mathbb{F}_2^k$ of the decoded codeword}

\BlankLine
Initialize: $\{\boldsymbol{v}^{(j)}\} \gets \emptyset$;

\For{$i = 0, 1, \dots, \ell_{\max} - 1$}{
    Generate a TEP  $e^{(i)}$;

    Compute test sequence:
    $\boldsymbol{a}^{(i)} = \boldsymbol{z} \oplus e^{(i)}$;

    Perform BM decoding on $\boldsymbol{a}^{(i)}$\;

    \If{BM decoding succeeds}{
    	obtain candidate codeword $\boldsymbol{v}^{(j)}$\;
    	Add $\boldsymbol{v}^{(j)}$ to $\{\boldsymbol{v}^{(j)}\}$;
    	
    	\If{$\boldsymbol{v}^{(j)}$ satisfies \eqref{eq:criterion2}}{
    		$\boldsymbol{v}_{\rm{best}} = \boldsymbol{v}^{(j)}$;
    		
    		\KwBreak\;
    	}
    }
}

\eIf{$\{\boldsymbol{v}^{(j)}\}\ne \emptyset$}{
	\If{$i = \ell_{\max}$}{
     Select best candidate by metric
    \begin{equation}
	{\boldsymbol{v}_{\rm{best}}}\mathop { = \arg \min }\limits_{\boldsymbol{v} \in \{ {\boldsymbol{v}^{(j)}}\} } \lambda (\boldsymbol{r},\boldsymbol{v});
    \end{equation}
}
Return information data  $\boldsymbol{u}$ corresponding to $\boldsymbol{v}_{\rm{best}}$;
}{
\Return{decoding failure};
}

\caption{ORB-Chase}
\end{algorithm}

The complete decoding algorithm is summarized in Algorithm 1. The traditional Chase decoding algorithm requires traversing  ${2^p}$ test patterns to find the optimal codeword. To achieve good performance, $p$ must be increased, which leads to an exponential growth in decoding complexity. Compared with the Chase algorithm, ORB-Chase can flexibly set the maximum number of test patterns to limit the growth of complexity.

\section{Numerical Results}
In this section, we present some simulation results highlighting the performance of the proposed ORB-Chase algorithm as well as the decoding complexity analysis. All simulations are conducted under the assumption of the BPSK signal transmission over an AWGN channel.

To compare the computational complexity of the Chase and ORB-Chase algorithms, we focus on the dominant decoding complexity after the initial reliability sorting. Since both algorithms perform the same initial sorting operation, this fixed preprocessing cost is excluded from the comparison. The remaining computational complexity is dominated by the number of BM decoding calls, whereas test error pattern generation, integer partition processing, and the proposed integer-based early termination criterion involve only simple integer operations and comparisons, whose computational overhead is negligible compared with BM decoding. Therefore, the average number of BM decoding calls is adopted as the primary complexity metric throughout this paper.

\begin{figure}\center	
	\includegraphics[width=8.0cm]{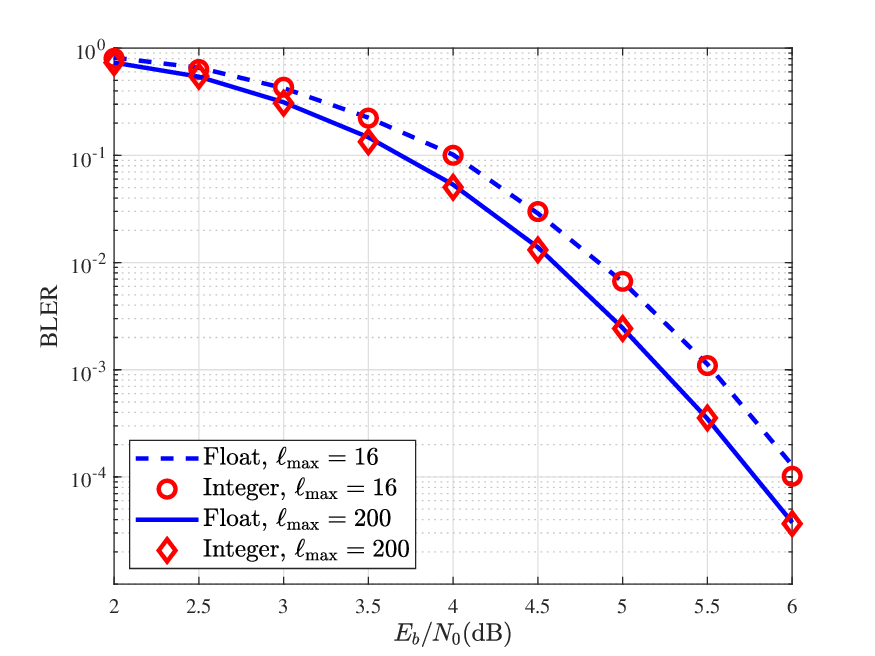}\\
	\caption{The BLER performance of the BCH code (127,113,5) under ORB-Chase decoding with the Optimality Criterion and the Integer-based Early Termination Criterion.}
	\label{criterion_BLER}
\end{figure}

\begin{figure}\center	
	\includegraphics[width=8.0cm]{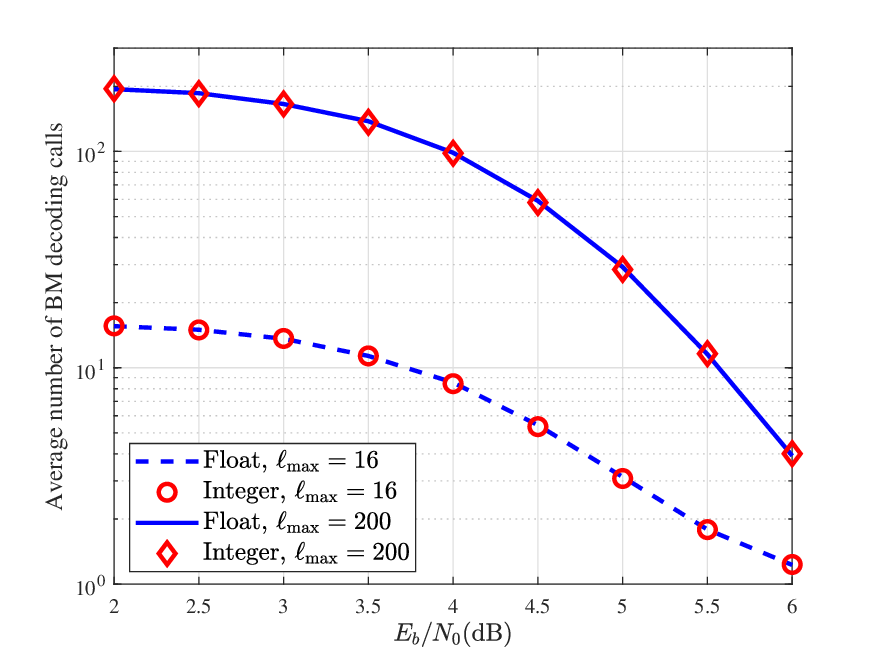}\\
	\caption{The average number of BM decoding calls for the BCH code (127,113,5) under ORB-Chase decoding with the Optimality Criterion and the Integer-based Early Termination Criterion.}
	\label{criterion_BM}
\end{figure}

\textbf{Example 1:} We first verify the effectiveness of the \emph{Integer-based Early Termination Criterion}. We simulate a BCH code (127,113,5) decoded with the ORB-Chase algorithm under the \emph{Optimality Criterion} and the \emph{Integer-based Early Termination Criterion}, where the maximum number of test patterns is \(\ell_{\text{max}}=\)16 or 200. The block error rate (BLER) performance curves under the two criteria are shown in Fig. \ref{criterion_BLER}. We can see that the BLER performance of the \emph{Integer-based Early Termination Criterion} almost coincides with that of the \emph{Optimality Criterion}, with no significant performance loss for different values of \(\ell_{\text{max}}\).
The average number of BM calls is shown in Fig. \ref{criterion_BM}, which indicates that the performance of the two criteria is almost identical.

The simulation results show that the proposed \emph{Integer-based Early Termination Criterion} makes the same early-termination decision as the \textit{Optimality Criterion} in almost all simulated cases. Therefore, although the proposed criterion is implemented using integer reliability values, it provides an accurate practical approximation for ML codeword identification while significantly reducing implementation complexity.

\begin{figure}\center	
	\includegraphics[width=8.0cm]{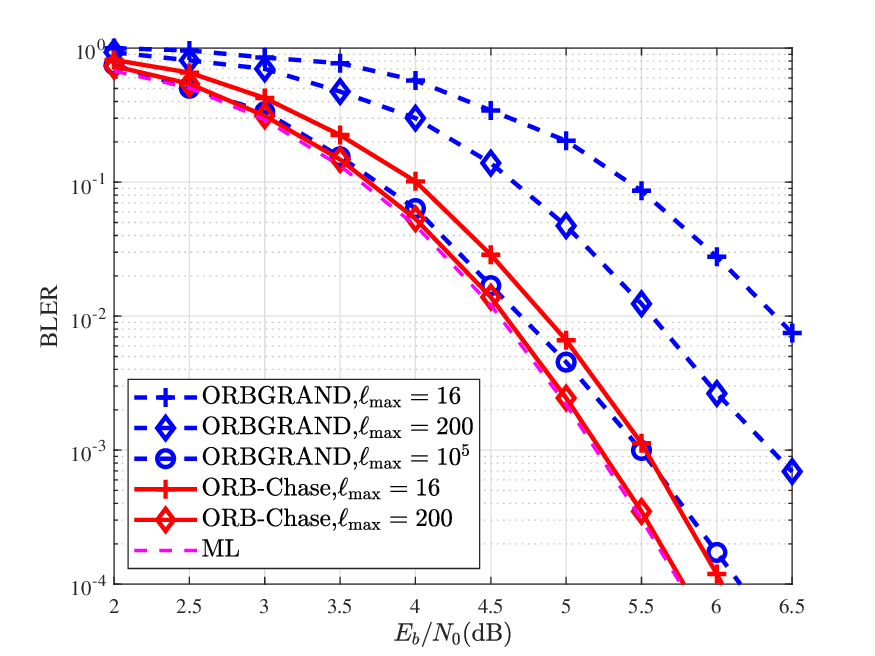}\\
	\caption{The BLER performance of the BCH code (127,113,5) under the ORBGRAND with \( \ell_{\text{max}} = 16, 200, 10^5\) and the ORB-Chase with \( \ell_{\text{max}} = 16, 200\).}
	\label{ORBGRAND}
\end{figure}
	
\textbf{Example 2:} We consider the BLER performance comparison of the BCH code with parameters (127,113,5) decoded with the ORB-Chase algorithm and the ORBGRAND algorithm, as shown in Fig. \ref{ORBGRAND}.
The ML decoding performance is obtained from \cite{Helmling}. For the ORBGRAND algorithm, the maximum number of codebook queries, $\ell_{\text{max}}$, is set to 16, 200, or \( 10^5 \). For the ORB-Chase algorithm, the maximum number of test patterns, $\ell_{\text{max}}$ , is set to 16 or 200.

At BLER = $10^{-3}$, for $ \ell_{\text{max}} = 16$ and $ \ell_{\text{max}} = 200$, the proposed ORB-Chase algorithm outperforms the ORBGRAND algorithm by about 1.5 dB and 1.0 dB, respectively. Furthermore, the ORBGRAND algorithm with $ \ell_{\text{max}} = 10^5$ performs similarly to the proposed ORB-Chase algorithm with $ \ell_{\text{max}} = 16$, while worse than the proposed ORB-Chase algorithm with $ \ell_{\text{max}} = 200$. We also observe that the proposed ORB-Chase algorithm with \( \ell_{\text{max}} = 200 \) approaches the ML lower bound for the considered BCH code, whereas the ORBGRAND with $ \ell_{\text{max}} = 10^5$ does not.		

\begin{figure}\center	
	\includegraphics[width=8.0cm]{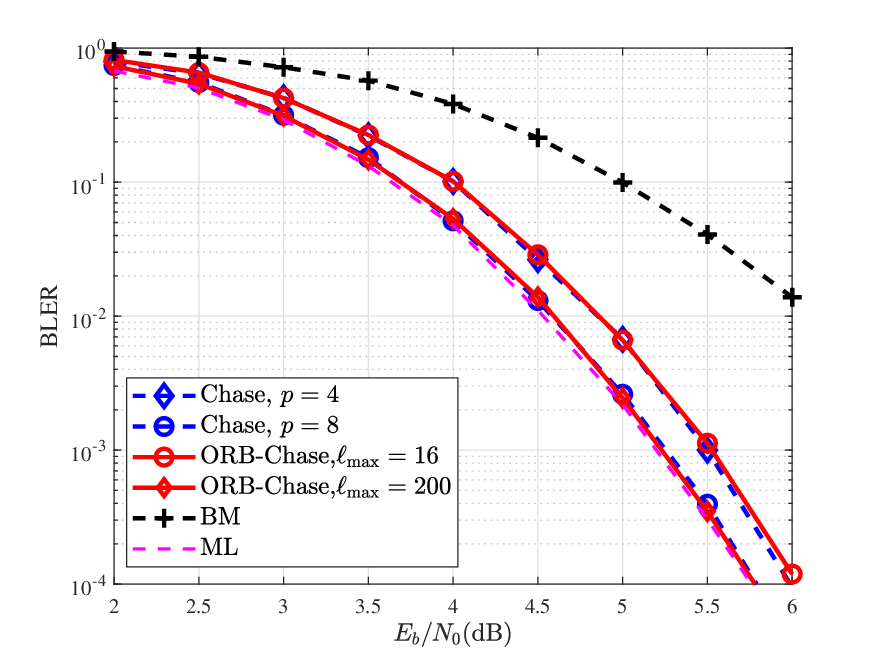}\\
	\caption{The BLER performance of the BCH code (127,113,5) under the Chase with \(p = 4, 8\) and the ORB-Chase with \(\ell_{\text{max}}=16, 200\).}
	\label{BCHa}
\end{figure}
\begin{figure}\center	
	\includegraphics[width=8.0cm]{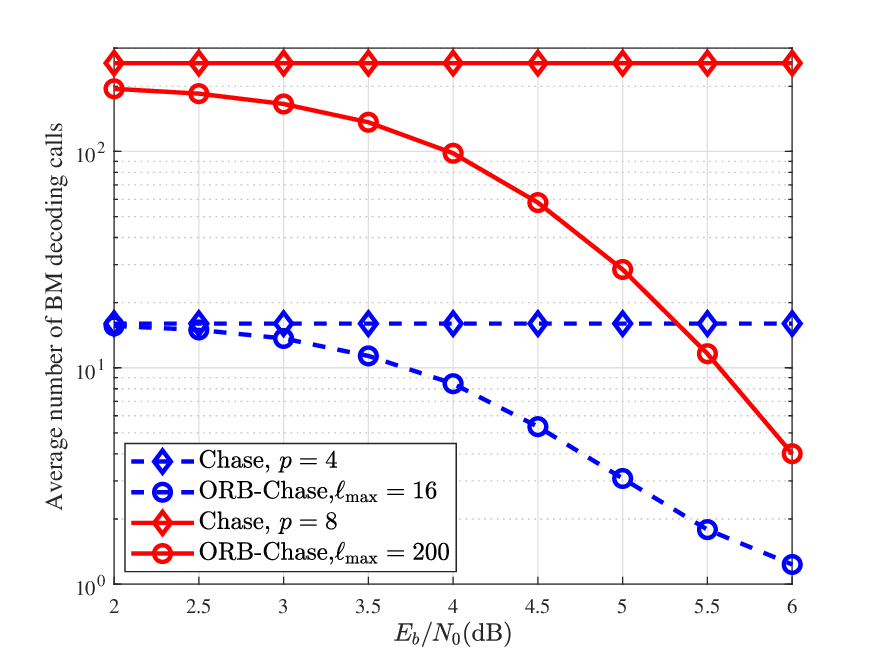}\\
	\caption{The average number of BM decoding calls for the BCH code (127,113,5) under the Chase with \(p = 4, 8\) and the ORB-Chase with \(\ell_{\text{max}}=16, 200\).}
	\label{BCHb}
\end{figure}
\textbf{Example 3:} We next consider a BCH code with parameters (127,113,5). To ensure a fair comparison, we configure each algorithm with two settings: the Chase algorithm with $p = 4$ and $p = 8$, corresponding to maximum numbers of test patterns of $2^4 = 16$ and $2^8 = 256$, respectively; and the ORB-Chase algorithm with $\ell_{\text{max}} = 16$ and $200$, respectively.
	
From Fig. \ref{BCHa}, we can see that the ORB-Chase algorithm requires a smaller \( \ell_{\text{max}} \) than the Chase algorithm to achieve the ML lower bound. The computational complexity comparison of the two algorithms under the same performance is depicted in Fig. \ref{BCHb}. As \(E_b / N_0 \) increases, the average number of BM decoding calls for the ORB-Chase algorithm decreases rapidly, requiring few BM decoding attempts on average at high \(E_b / N_0 \). Compared with the Chase algorithm achieving the same performance, the average number of BM decoding calls of the ORB-Chase algorithm with \(\ell_{\text{max}} = 16 \) and 200 is reduced by 47.2\% and 50.9\%, respectively, at \(E_b / N_0 = 4\,\text{dB} \), and by 92.3\% and 98.0\%, respectively, at \(E_b / N_0 = 6\,\text{dB} \).
\begin{figure}\center	
	\includegraphics[width=8.0cm]{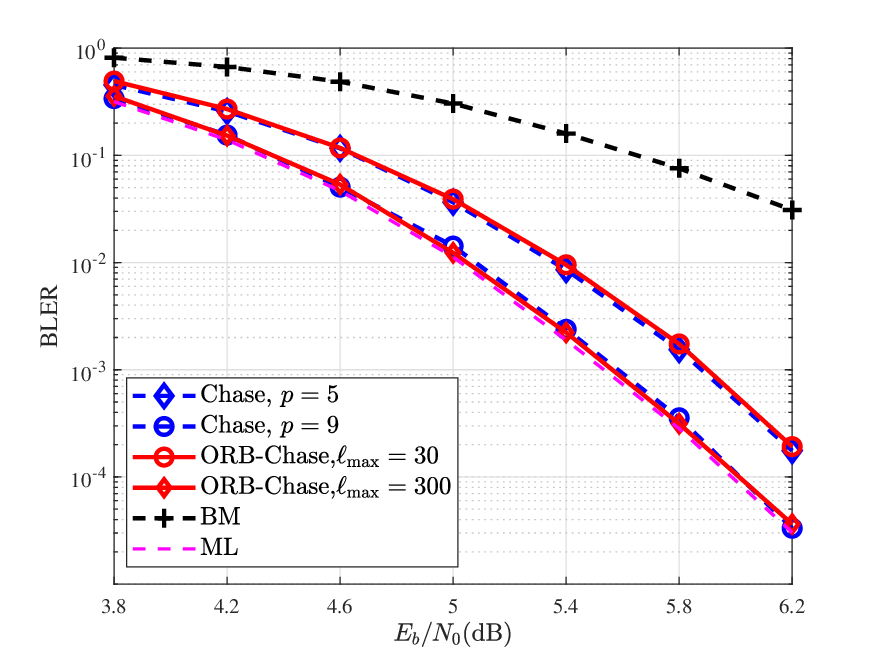}\\
	\caption{The BLER performance of eBCH code (256,239,6) under the Chase with \( p = 5, 9\) and the ORB-Chase with \(\ell_{\text{max}}=30, 300\).}
	\label{eBCHa}
\end{figure}
\begin{figure}\center	
	\includegraphics[width=8.0cm]{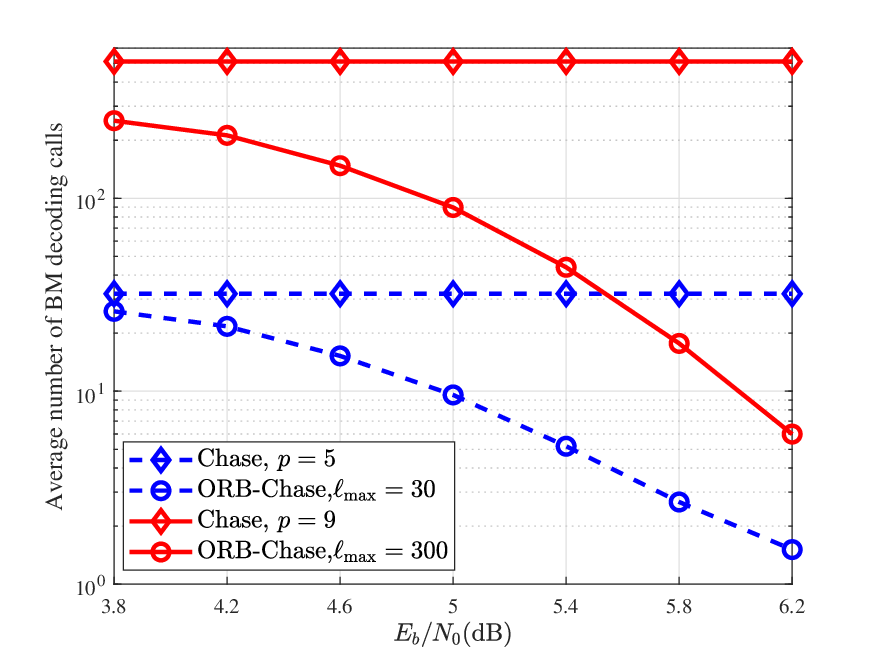}\\
	\caption{The average number of BM decoding calls for the eBCH code (256,239,6) under the Chase with \( p = 5, 9\) and the ORB-Chase with \(\ell_{\text{max}}=30, 300\).}
	\label{eBCHb}
\end{figure}

\textbf{Example 4:} We also consider an eBCH code with parameters (256,239,6). We simulated two decoding algorithms: the Chase algorithm with
\( p = 5 \) and \( p = 9 \), and the ORB-Chase algorithm with \(\ell_{\text{max}} = 30 \) and \(\ell_{\text{max}} = 300 \).
	
Fig. \ref{eBCHa} shows that the ORB‑Chase algorithm can achieve performance identical to that of the Chase algorithm while requiring a smaller \(\ell_{\text{max}}\), especially near the ML bound.
In Fig. \ref{eBCHb},  a complexity comparison of the two algorithms under the same performance is given. Compared with the Chase algorithm at the same BLER performance, the average number of BM decoding calls of the ORB-Chase algorithm with \( \ell_{\text{max}} = 30 \) and 300 is reduced by 49.2\% and 50.8\%, respectively, at \( E_b / N_0 = 4.6\,\text{dB} \), and by 94.9\% and 98.1\%, respectively, at \( E_b / N_0 = 6.2\,\text{dB} \). As \( E_b / N_0 \) increases, the average number of BM decoding calls for the ORB-Chase algorithm decreases rapidly, requiring only a few BM decoding attempts in the high SNR region, demonstrating a significant complexity advantage.
These results verify that the proposed algorithm can substantially reduce computational complexity while maintaining good error correction performance.

\section{Conclusion}
In this paper, we proposed an ordered-reliability-bits Chase (ORB-Chase) decoding algorithm for BCH codes, which achieves near-optimal performance with significantly reduced complexity compared to the traditional Chase algorithm. Simulation results for BCH codes with parameters (127,113,5) show that the proposed ORB-Chase algorithm outperforms the ORBGRAND under a limited number of searches. Moreover, simulations shows that the average number of BM decoding calls decreases rapidly as \( E_b / N_0\) increases, allowing the ORB-Chase decoder to achieve the same performance as the Chase algorithm with substantially fewer BM operations. These results confirm that the ORB-Chase algorithm offers an excellent trade-off between error correction performance and computational complexity, making it a promising candidate for practical applications.

In future work, we will consider optimizing the generation order of test error patterns, such as the improved logical weight ordering in \cite{Condo2021}, to further reduce decoding complexity.

\end{document}